\documentclass{ws-ijmpd}

\begin{document}

\markboth{Zhen Cao}
{The ARGO-YBJ Experiment Progresses and Future Extension}

\catchline{}{}{}{}{}
%
%%%%%%%%%%%%%%%%%%%%%%%%%%%%%%%%%%%%%%%%%%%%%%%%%%%%%%%%%%%%%%%%%%%%

\title{The ARGO-YBJ Experiment Progresses and Future Extension\footnote{A proceeding of the 1st Galileo-Xu Guangqi Meeting at Shanghai, Oct 26-30, 2009}  }

\author{Zhen Cao\footnote{On behalf of the ARGO-YBJ Collaboration and the LHAASO Collaboration}}

\address{Institute of High Energy Physics\\
19B Yuquan St\\
Shijingshan, Beijing 100049,
China\\
caozh@ihep.ac.cn}

\maketitle

\begin{history}
\received{Day Month Year}
\revised{Day Month Year}
\comby{Managing Editor}
\end{history}

\begin{abstract}
Gamma ray source detection above 30TeV is an encouraging approach for finding galactic cosmic ray origins. All sky survey for gamma ray sources using wide field of view detector is essential for population accumulation for various types of sources above 100GeV. To target the goals, the ARGO-YBJ experiment has been established. Significant progresses have been made in the experiment. A large air shower detector array in an area of 1km$^2$ is proposed to boost the sensitivity. Hybrid detection with multi-techniques will allow a good discrimination between different types of primary particles, including photons and protons, thus enable an energy spectrum measurement for individual specie. Fluorescence light detector array will extend the spectrum measurement above 100PeV where the second knee is located. An energy scale determined by balloon experiments at 10TeV will be propagated to ultra high energy cosmic ray experiments
\end{abstract}


\section{Introduction}	

More than 100 sources\cite{sources} with emission of gamma rays around 1 TeV have been discovered\cite{HESS-1} in the last two decades. The mechanism of the emission has been investigated somewhat thoroughly among the discovered sources. Evidences\cite{HESS-2} support a scenario in which all kinds of soft photons such as star light or cosmic microwave background (CMB) photons near an object that is associated with a strong shock wave can be converted into TeV gamma rays because they may be knocked by high energy electrons accelerated at the front of the shock waves.  Pulsar wind nebulae (PWN), supernova remnants (SNR), X-ray binaries (Micro-quasars) and active galactic nuclei (AGN) are main categories of gamma ray sources.  Energy spectrum distributions are measured up to nearly 100TeV for some of the sources\cite{Felix}. Most of the observed energy spectra are interpolated quite well using the mechanism based on the inverse Compton scattering model, however, some of them seem to be difficult to fit, thus a more interesting mechanism of gamma ray production based on neutral pion decay is introduced by many authors\cite{pi0}. It implied that the sources with such a feature may be origins of cosmic rays because pions must be produced in collisions between accelerated protons or nuclei (cosmic ray particles) and ambient material. For further investigation, a dedicated detector with very high sensitivity at high energies above 30TeV is required for a clean measurement of gamma ray energy spectrum without any contemplation from cosmic rays. Due to absorption of extra-galactic background light(EBL) or CMB, gamma rays at such high energy can not propagate through the space between galaxies, in other words, the sources must be inside of our galaxy. 
There exist even more complicated phenomena, such as strong transient behaviors of AGNs or gamma ray bursts, etc. the flux of gamma rays could change by orders of magnitudes within a few hours. In order to understand what causes such strong and quite frequently flares, collecting sufficient population for different types of sources is necessary. Lots of interesting physics, such as quantum gravitational effects etc, can be studied through the flare observation. 
Narrow field of view (FOV) Cherenkov telescopes for pointing investigation have been very successful in finding most of point sources including highly variable AGNs. With an intrinsic difficulty, namely not being able to operate in cloudy weather or with the moon in the sky, the Cherenkov telescopes are not optimized towards monitoring the transient phenomena. An all-sky survey for gamma ray sources will be an essential approach. A ground based large air shower detector array at high altitude is a natural response to the strong demand. The ARGO-YBJ experiment is a successful approach along the direction. In this paper, the progresses of the experiment are briefly summarized in the next section. In order to boost the discovery power of the wide FOV detector in the all sky survey, a much larger detector array is designed as a major upgrade of the current experiment. A layout of the new array and corresponding design of detectors are presented in the following sections. Expected performance of the new array is also reported.

Such a detector array is not only useful for the gamma ray source survey but also plays an important role of bridging between direct measurements of energy spectra of individual cosmic ray species at balloon heights and ultra high energy cosmic ray experiments on the ground, such as Auger and Telescope Array experiments. Matching with direct measurements that use calorimeters and charge sensitive detectors in 10TeV regime sets an absolute energy scale for air shower experiments on the ground. The ARGO-YBJ experiment plays an essential role in the measurement of the spectrum in the low energy domain. A hybrid of multiple detection techniques plays increasingly important role in the future experiments. As a major upgrade, the future experiment at the high altitude of 4300m a.s.l. is designed to cover a wide energy region from 30TeV to few EeV by combining multiple air shower detection techniques together. The layout of the detector array and corresponding performance are described in the sections before concluding the paper.

\section{The ARGO-YBJ Experiment and Its Results}
The ARGO-YBJ experiment, located at the YangBaJing Cosmic Ray
 observatory (Tibet, China, $4300$m a.s.l.), is a single layer of
 Resistive Plate Chambers (RPCs)\cite{argo1} on a surface
 of 78 $\times$ 74m$^{2}$. The central
 part has a full coverage area of 5800m$^2$ with 10$\times$13
 clusters which have 12 RPCs in each and each RPC has
10 pads as the basic unit of the detector. The central carpet array is used for event triggering. Outside the carpet,  a skirt part of the array,
so-called guard ring, made of 23 clusters with a coverage of 25\%, 
is designed to select events that cores of showers are falling
inside the array. Cosmic ray data have been collected with a trigger
criterion of
 $N_{hit}\geq 20$, where N$_{hit}$ is number of fired pads in a
 trigger window of 420 ns. The trigger rate for
 air shower events is $\sim$3.6kHz. Cosmic ray data from July 2006 to
 October 2009 are analyzed. The overall exposure is 825.25 days with a duty circle greater than 85\% 
since October 2007. 
With a measurement of a shadow of galactic cosmic rays blocked by the moon, one learns the angular resolution for the shower arrival direction is better than 0.5$^\circ$ for showers that $N_{hit}>100$. The significance of the shadow is 55$\sigma$. This indicates that the ARGO-YBJ detector is running very stable and steadily collecting cosmic ray events.
\begin{figure}[pb]
\centerline{\psfig{file=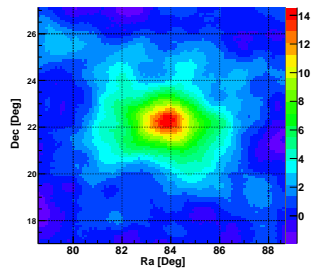,width=4.1cm}}
%\vspace*{8pt}
\caption{Significance map of the Crab nebula using the ARGO-YBJ data.\label{Crab}}
\end{figure}

\begin{figure}[pb]
\centerline{\psfig{file=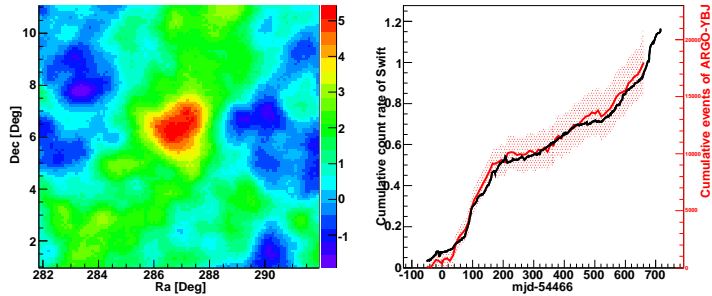,width=9.7cm}}
%\vspace*{8pt}
\caption{Left: Significance map of the gamma ray source MGRO1908+06 discovered by the Milagro experiment. Right: Accumulative excesses from Mrk421 direction correlated with the hard X-ray flux measured by the satellite based Swift telescope. Black curve represents the hard X-ray flux and the red dots with error bars are results from ARGO-YBJ.\label{MGRO_Mrk}}
\end{figure}

To estimate the sensitivity of the ARGO-YBJ detector in terms of its accumulative exposure, the Crab nebula as a standard candle for $\gamma$ ray sources in the very high energy region is measured. Photons from the nebula produce an excess of 14.5$\sigma$ at the maximum over the cosmic ray backgrounds, as shown in Fig.\ref{Crab}. With such a sensitivity, many gamma ray astronomic investigations are able to be performed, such as a survey in the whole sky for brightest sources and a monitor for transient effects associated with variable sources, e.g. AGN flares and gamma ray bursts.  
As examples of a steady source in our galaxy and a blazer that bursts in flares occasionally, our measurement of the newly found MGRO1908+06\cite{Milagro_1} is shown in the left panel in Fig.\ref{MGRO_Mrk} with a significance of 5.4$\sigma$. Mrk421 has many flares in the past years and is also observed by ARGO-YBJ with 11.5$\sigma$. Most importantly, the observation of the highly active blazer in a long run demonstrates the power of monitoring for those types of transient phenomena using ARGO-YBJ detector. In the right panel in Fig.\ref{MGRO_Mrk}, an accumulation of the excesses in the Mrk421 direction reveals a nice correlation with the hard X-ray (15-50keV) flux measured by the space borne X-ray telescope\cite{Swift}.
\begin{figure}[pb]
\centerline{\psfig{file=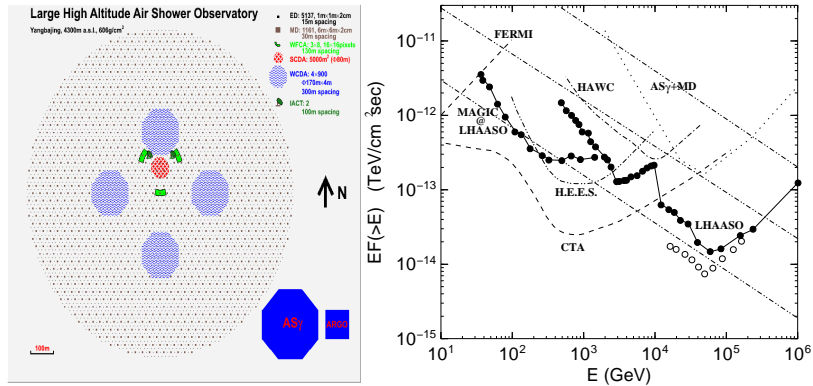,width=10.7cm}}
%\vspace*{8pt}
\caption{Left: A layout of the LHAASO array. Right: The sensitivity of the major experiments and future projects for gamma ray astronomy，solid dots for LHAASO, dashed for CTA, dash-dotted for HAWC, dash-double-dotted for AS$_\gamma$+MD and dotted for existing experiment HEES as a comparison. \label{array}}
\end{figure}

\section{A Design of Detector Array for LHAASO}
In order to fulfill all the goals mentioned in Sec.1, a large scale complex of many kinds of detectors is needed beyond the ARGO-YBJ experiment. For gamma ray source surveys, a water Cherenkov detector array(WCDA) with a total active area of 90,000m$^2$ is proposed, marked by the four octagons in the left penal in Fig.\ref{array} It is sensitive to gamma ray showers above a few hundred GeV and will achieve a sensitivity of detecting sources that have emission intensity of at least 2\% 
of the emission from the Crab Nebula I$_{crab}$ with a significance of 5$\sigma$ per year, as shown in the right penal in Fig.\ref{array}. For discovered sources in the surveys, two types of further investigations can be pursued. 

One is to measure the energy spectra of gamma rays up to few hundreds of TeV for searching for galactic cosmic ray origins. The focus is the high energy ends of the spectra where one expects to see differences due to different origins of gamma rays, either from inverse Compton scattering of high energy electrons or from decays of neutral pions which are produced in high energy proton interaction with ambient material near the sources. For this purpose, a particle detector array with an effective area of 1km$^2$ is proposed (KM2A) including a muon detector array with 40,000m$^2$ active area. This allows a clean measurement of gamma ray spectra above 60TeV without any hadronic shower background by selecting muon-less showers. 5137 scintillator detectors (1m$^2$ each), shown as the smallest dots in the left penal in Fig.\ref{array}, are used to measure arrival directions and total energies of showers. 1200 $\mu$ detectors made of 36 scintillator detectors same as electron detectors but covered by absorber such as 2.5m dirt or 5m water, shown as squares in the left penal in Fig.\ref{array}, are used for muon content measurements. An expected sensitivity of this array is also plotted in the right penal in Fig.\ref{array} connecting with the WCDA sensitivity at ~10TeV. It makes a perfect complementary to the narrow field view CTA experiment which is the most sensitive detector at lower energies and details a morphologic probe within a FOV of ~3$^\circ$ around the sources in the next 5 years.

The other is to carry out similar morphologic probes at much lower energies, e.g. ~30GeV as a further extension of the project. It requires adding two MAGIC-II type large size telescopes (LST), similar to that will be used in the central part of CTA, to the LHAASO project at an altitude of 4300m a.s.l., as marked by solid dots in the  penal in Fig.\ref{array}. 

To measure cosmic ray spectra with individual composition, a multi-parameter investigation is necessary. In general, the shower maximum, the muon content and the high energy component near the core are three independent parameters that can be used for deducing signatures of different showers induced by different nuclei. Two more detector components are proposed for measuring those parameters except for the muon content. They are the 24 wide FOV Cherenkov telescope array (WFCA) and the high threshold core-detector array (SCDA) with an effective area of 5000m$^2$, shown as three rectangles and a very dense area at the center of the array in the left penal in Fig.\ref{array}, respectively.

\section{Physics Perspectives of the LHAASO Project}
\subsection{Investigation for cosmic ray origins among galactic gamma ray sources}
RXJ1713-3946, one of the brightest gamma ray sources in the southern sky, has been thoroughly investigated\cite{pi0}  in terms of spectrum measurements and very detailed morphologic probing. As a shell type SNR, it has been naturally considered as a candidate of a cosmic ray source, moreover
it is difficult to be interpolated as a source with a pure electron origin model according to its gamma ray spectrum. However, it has yet been so certain that this SNR is a source of cosmic ray protons. A concrete evidence for existence of accelerated protons relies on an accurate measurement of the spectrum extended to higher energies and even more important a collection of many similar sources in our galaxy that have the same feature. Using LHAASO KM2A, one will carry out a background-free measurement of spectra for such sources above 60TeV. All spectra of discovered sources by HESS are well above the LHAASO sensitivity below few hundred TeV if those sources appear in the northern sky. More than 30 events above highest energies ever measured are expected for most of the sources in 3 years. According to recent theoretical investigation\cite{ZhangLi}, old SNRs have greater chances to have a capability of accelerating protons to high energies, thus to produce higher energy photons than the inverse-Compton scattering of electrons. All the physics goals set a minimum of requirements on the LHAASO KM2A performance, such as an angular resolution of the array better than 0.5$^\circ$ and an energy resolution of the array better than 20\%,
which are all tested successfully in the existing experiments at the Tibet site.

\subsection{Full-sky survey for the collection of a gamma ray source population}
The number of sources with TeV gamma ray emission has been exponentially increasing since the discovery of the first source, the Crab Nebula, using a narrow FOV Cherenkov telescope in 1989. Without a guide from surveying results, an efficiency of the discovery of TeV gamma sources remains at 
10\% 
among very carefully selected candidates. After a strong progress in the last twenty years, the rate seems to start slowing down. According to experience in optical and X-ray astronomy, if there is no boost in sensitivity of the observational survey, the population of sources will not be expected increasing with the same pace. A demand in the community to have the observational survey with equal sensitivity as the narrow FOV observation, becomes stronger recently. Ground based shower detector arrays, such as the Milagro experiment, start to demonstrate their power of discovery of sources, especially to sources with spatial extension which is particularly difficult to be observed by the narrow FOV observation. Following Milagro’s recent results\cite{Milagro} of confirming some sources found by the space borne FERMI/LAT detector, the ARGO-YBJ experiment confirmed as well that adding all sources in a category together, such as all brightest pulsar TeV sources, excesses are observed given a statistics collected in an operation for slightly longer than 2 years. The LHAASO project will have a sensitivity of seeing all sources in the whole northern sky that are stronger than 0.02$I_{crab}$ simultaneously. With a similar sensitivity, HESS has surveyed a very small region near the center of our galaxy that resulted in discoveries of numerous sources. At least 99\% 
of the sky is yet to be surveyed. With an operation of LHAASO for 1$\sim$2 years, one half of the sky will be surveyed just like the central region of our galaxy.

\subsection{``Knees" of individual species and an absolute energy scale}
A detector array like LHAASO at a height of 4300m a.s.l. must be used for cosmic ray research because showers around the ``knee" just reach their maximum at 1$\sim$2km above the array, thus effects due to shower fluctuations are minimized. Since the ``knee" was found, its origin has yet to be clarified. A major difficulty is how to separate different primary species from each others in shower observations. Accurately measuring muon contents in showers with a large enough active detector array is one of the handles. LHAASO array, equipped with the largest muon detector array ever, should make its contribution to this topic. However, many historic experiments, such as CASA/MIA and Kascade, demonstrated that it is not sufficient to effectively separate species by using the muon content only. With a small portion of extension by adding WFCA and SCDA to the project, shower maximum depths, X$_{max}$ and high energy fluxes carried by particles near the shower cores can be measured. Simulation shows that the X$_{max}$ can be determined with a resolution of 50g/cm$^2$ which can be useful for the separation, while a difference of 150g/cm$^2$ between protons and irons in average is expected in this energy range. Simulation also shows that SCDA will help to separate proton showers from others with high purity and efficiency. Working together, the three independent pieces of information will greatly enhance the selecting power for protons, helium and iron nuclei on an event by event bases above 30TeV. This is particularly important because the spectra of those particles had been measured directly in numerous balloon borne experiments that set an absolute energy scale for the air shower observation. Using an array of 5000m$^2$, the spectra will be extended up to 10PeV or even higher within two or three years. 

\subsection{Extension to UHE for the second ``knee"}
\begin{figure}[pb]
\centerline{\psfig{file=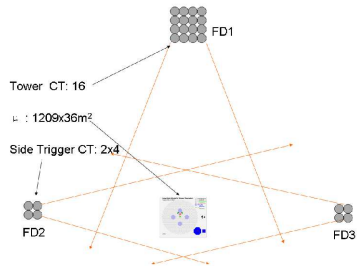,width=4.7cm}}
%\vspace*{8pt}
\caption{Layout of the fluorescence detector array and the LHAASO array\label{tower}}
\end{figure}

To extend the spectrum to higher energies and make a connection with experiments, such as TA and Auger, at altitudes around 1600m a.s.l., the wide FOV telescopes will be re-arranged to measure shower fluorescence light and monitoring the space above the ground array from a distance of 4$\sim$5km. The detector configuration is shown in Fig.\ref{tower}, in which the main detector array is composed of 16 telescopes covering elevations from 3$^\circ$ to 59$^\circ$ and two other detector arrays, covering elevations from 3$^\circ$ to 31$^\circ$, to observe showers from perpendicular directions. Showers above 100PeV will be detected stereoscopically to maintain a high resolution of X$_{max}$. Combining with the muon content measured by KM2A, the telescopes will achieve the spectrum and the composition measurements around the second knee.

\section{Conclusion}
The ARGO-YBJ experiment has demonstrated a big potential in whole sky survey observation for both source population accumulation and transient phenomenon monitoring with an operation for 2$\sim$3 years. 
The LHAASO project is designed for greatly boost the capability, thus gamma ray sources will be surveyed above a few hundreds GeV. Complementary with the CTA project, the focus of LHAASO experiment is mainly on full-sky survey and galactic cosmic ray origin search above 30TeV. With an extension of using Cherenkov telescopes designed by the Magic group, a detailed source morphologic investigation is also in the extended scope of the project. To maximize the advantage of being at high altitude, cosmic ray spectrum and composition will be measured over a wide energy range spanning a bridge between the balloon borne measurements for each species and UHECR observations above 1EeV at lower altitudes.

\section*{Acknowledgments}
This work is supported by Knowledge Innovation fund (U-526) of IHEP, China.

%\begin{thebibliography}{000} %for 3 digits
%\begin{thebibliography}{00}  %for 2 digits

\end{document}